\documentclass[prd,nofootinbib,preprint,superscriptaddress,twocolumn,10pt]{revtex4}
\pdfoutput=1
\usepackage{amsmath,amssymb}
\usepackage{epsfig}
\usepackage{graphicx}
\usepackage[usenames,dvipsnames]{color}
\usepackage{subfigure}
\usepackage{slashed}
\usepackage{comment}
\usepackage[normalem]{ulem}
\usepackage[colorlinks,citecolor=blue]{hyperref}
\usepackage{color}

\usepackage{multirow}

\usepackage{tikz-feynman}
\tikzfeynmanset{compat=1.1.0}

\usepackage{tikzsymbols}

\usepackage{xcolor}
\usepackage{soul}

\begin{document}
\title{Large Neutrino Asymmetry from TeV Scale Leptogenesis}

\author{Debasish Borah}
\email{dborah@iitg.ac.in}
\affiliation{Department of Physics, Indian Institute of Technology Guwahati, Assam 781039, India}

\author{Arnab Dasgupta}
\email{arnabdasgupta@pitt.edu}
\affiliation{Pittsburgh Particle Physics, Astrophysics, and Cosmology Center, Department of Physics and Astronomy, University of Pittsburgh, Pittsburgh, PA 15206, USA}

\begin{abstract}
We study a class of leptogenesis scenarios with decay or scattering being the source of lepton asymmetry, which can not only give rise to the observed baryon asymmetry in the universe but also can leave behind a large remnant neutrino asymmetry. Such large neutrino asymmetry can not only be probed at future cosmic microwave background (CMB) experiments but is also motivating due to its possible role in solving the recently reported anomalies in $^4{\rm He}$ measurements. Additionally, such large neutrino asymmetry also offers the possibility of cogenesis if dark matter is in the form of a sterile neutrino resonantly produced in the early universe via Shi-Fuller mechanism. Considering $1 \rightarrow 2, 1 \rightarrow 3$ as well as $2 \rightarrow 2$ processes to be responsible for generating the asymmetries, we show that only TeV scale leptogenesis preferably of $1 \rightarrow N \, (N \geq 3)$ type can generate the required lepton asymmetry around sphaleron temperature while also generating a large neutrino asymmetry $\sim \mathcal{O}(10^{-2})$ by the epoch of the big bang nucleosynthesis. While such low scale leptogenesis can have tantalising detection prospects at laboratory experiments, the indication of a large neutrino asymmetry provides a complementary indirect signature.
\end{abstract}

\maketitle
\noindent
\textbf{\textit{Introduction:}} 
The baryonic matter content in the present universe is highly asymmetric leading to the longstanding puzzle of baryon asymmetry of the universe (BAU). This observed excess of baryons over anti-baryons is quantified in terms of the baryon to photon ratio as \cite{Planck:2018vyg} 
\begin{equation}
\eta_B = \frac{n_{B}-n_{\overline{B}}}{n_{\gamma}} \simeq 6.2 \times 10^{-10}, 
\label{etaBobs}
\end{equation} 
based on the cosmic microwave background (CMB) measurements which also agrees well with the big bang nucleosynthesis (BBN) estimates \cite{Zyla:2020zbs}. In order to generate the observed BAU dynamically, the Sakharov's conditions \cite{Sakharov:1967dj} are required to be satisfied which the standard model (SM) of particle physics fails to do in required amount. One appealing way to achieve baryogenesis is the leptogenesis \cite{Fukugita:1986hr} route where a non-zero lepton asymmetry is first generated which later gets converted into the BAU via electroweak sphalerons \cite{Kuzmin:1985mm}. While the observational constraints related to BAU restricts the net lepton asymmetry around sphaleron decoupling temperature $(T_{\rm sph})$, it is possible to generate large lepton asymmetry at lower temperatures $(T < T_{\rm sph})$ while being consistent with the observed BAU. However, charge neutrality of the early universe restricts the asymmetry in charged lepton sector to be at most of the order of $\eta_B$. This leaves us with the only option of storing large lepton asymmetry in the neutrino sector. Interestingly, such large neutrino asymmetry can be probed experimentally via precision measurements of relativistic degrees of freedom $N_{\rm eff}$ at CMB experiments. Additionally, it can also affect BBN estimates, as pointed out recently in the light of anomalous observations related to primordial Helium-4 ($^4{\rm He}$) abundance.

The recent near-infrared observation of 10 extremely
metal-poor galaxies by the Subaru Survey \cite{Matsumoto:2022tlr}, along with 54 previously observed galaxies, have led to the determination of the primordial abundance of $^4{\rm He}$ as $Y_P=0.2379^{+0.0031}_{-0.0030}$. While this is slightly smaller than earlier estimates \cite{2020ApJ, Aver:2015iza, Izotov:2014fga}, inclusion of the primordial deuterium constraints lead to a $>2\sigma$ tension between the predicted number of neutrino species $N_{\rm eff}=2.41^{+0.19}_{-0.21} $ and the standard model expectation $N_{\rm eff}=3.046$, referred to as the Helium anomaly \cite{Matsumoto:2022tlr}. Allowing a large neutrino asymmetry, quantified in terms of the degeneracy parameter of the electron type neutrino $\zeta_e = 0.05^{+0.03}_{-0.03}$\footnote{Neutrino asymmetry parameter is related to $N_{\rm eff}$ as $\Delta N_{\rm eff} = \frac{30}{7\pi^2} \sum_\alpha \zeta^2_\alpha$. The total neutrino asymmetry is defined as $ \eta_{\Delta L_\nu}= (n_\nu - n_{\overline{\nu}})/n_\gamma= \frac{\pi^2}{33 \zeta(3)} \sum_\alpha \zeta_\alpha$. Here $\alpha = e, \mu, \tau$.}, it is however possible to obtain a large $N_{\rm eff} = 3.22^{+0.33}_{-0.30}$ consistent with SM prediction within $1\sigma$ \cite{Matsumoto:2022tlr}. A more recent analysis of the BBN data jointly with information from the CMB observations also found evidence for a large neutrino asymmetry in the early universe at $\sim 2\sigma$ confidence level \cite{Burns:2022hkq}. While we consider large neutrino asymmetry solution to this Helium anomaly, there exists other solution too as considered in \cite{Kohri:2022vst} proposing a modified gravity origin. 

While these indications are only suggestive at this stage, it is tantalising to consider beyond standard model (BSM) scenarios which can create such large neutrino asymmetries. Even if the Helium anomaly disappears with future observations, large neutrino asymmetry can be probed experimentally at CMB experiments. Also, from dark matter (DM) model building point of view, such large neutrino asymmetry allows resonant production of sterile neutrino DM via Shi-Fuller mechanism \cite{Shi:1998km} which is relatively less constrained from X-ray bounds compared to the production via Dodelson-Widrow mechanism \cite{Dodelson:1993je}. One may refer to a review \cite{Drewes:2016upu} for details of these mechanisms and relevant experimental constraints. Among the BSM scenarios to generate large neutrino asymmetry, one recent attempt \cite{Kawasaki:2022hvx} along with a few related earlier works \cite{March-Russell:1999hpw, McDonald:1999in, Casas:1997gx} have utilised the Affleck-Dine mechanism \cite{Affleck:1984fy}. However, no studies have been done to relate it to conventional leptogenesis scenarios, from out-of-equilibrium decay or scattering of heavy particles. We, for the first time, point out that even in out-of-equilibrium decay leptogenesis, it is possible to generate such large neutrino asymmetry at late epochs while producing the required lepton asymmetry at sphaleron temperature $(T_{\rm sph})$ consistent with the observed baryon asymmetry. Since there exists only a single source of lepton asymmetry in such minimal setups, one requires the yield in lepton asymmetry to continue over a long period $T \in ( T_{\rm sph}, T_{\rm BBN})$. This also ensures that the small lepton asymmetry required for successful baryogenesis via leptogenesis \cite{Fukugita:1986hr} is produced by $T_{\rm sph}$ which later gets enhanced to a large neutrino asymmetry $\sim \mathcal{O}(10^{-2})$ by $T_{\rm BBN}$\footnote{While BBN occurs over a range of temperature below 10 MeV, we consider $T_{\rm BBN}=10$ MeV in order to ensure that the large neutrino asymmetry is produced before the onset of BBN.}. While it is possible to create a large neutrino asymmetry in electron neutrino type at high scale and be consistent with the observed baryon asymmetry by considering cancellation among lepton asymmetries of different lepton flavours \cite{March-Russell:1999hpw}, we do not consider such a fine-tuned scenario here. The requirement of a prolonged yield in lepton asymmetry even below sphaleron temperature requires the scale of leptogenesis to be low and also the mother particle to be out-of-equilibrium in order to avoid Boltzmann suppression in its number density.

Starting with a model-independent approach, we first consider decays of heavy Majorana fermions or right handed neutrinos (RHN) as origin of lepton asymmetry by taking the mass of RHNs, their decay widths and the CP asymmetry parameter as free parameters. We then show that it is not possible to fulfill both the criteria mentioned above with this simple vanilla leptogenesis setup \cite{Fukugita:1986hr, Davidson:2008bu}, even with resonantly enhanced CP asymmetry parameter \cite{Pilaftsis:2003gt}. We then consider the RHNs to have additional scattering processes responsible for keeping them in equilibrium at early epochs followed by late freeze-out, a characteristic feature of (but not limited to) three-body decay origin of leptogenesis \cite{Masiero:1992bv, Adhikari:1996mc, Sarkar:1996sn, Hambye:2001eu, Dasgupta:2019lha, Abdallah:2019tij, Grossman:2003jv, DAmbrosio:2003nfv, Fong:2011yx, Borah:2020ivi}. We also consider the possibility of leptogenesis from scattering, similar to the WIMPy leptogenesis \cite{Kumar:2013uca, Racker:2014uga, Dasgupta:2016odo, Borah:2018uci, Borah:2019epq, Dasgupta:2019lha} and find it to be unsuccessful in generating the required asymmetries. We show that TeV scale leptogenesis of $1 \rightarrow 2$ type decay having additional interactions responsible for producing the mother particle in equilibrium can, in principle, generate such large neutrino asymmetry with resonantly enhanced CP asymmetry while being consistent with the observer baryon asymmetry. On the other hand, in leptogenesis of $1 \rightarrow 3$ type decay which naturally leads to additional interactions keeping the mother particle in equilibrium in early universe, the desired asymmetries can be generated without any additional ingredients. Finally, we propose a concrete model for three-body decay leptogenesis consistent with successful leptogenesis, large neutrino asymmetry as well as light neutrino masses. While we do not discuss dark matter (DM) phenomenology in this work, such large neutrino asymmetry can lead to resonant production of sterile neutrino DM \cite{Shi:1998km, Abazajian:2001nj, Drewes:2016upu}. In fact, a recent work \cite{Eijima:2020shs} has shown that such large neutrino asymmetry required for resonant DM production can be generated in a minimal setup where oscillation is the primary source of asymmetry\footnote{See \cite{Foot:1995qk} for earlier work on generating large neutrino asymmetry from oscillation.}. Thus, our proposal also offers a DM-baryon cogenesis setup in the context of TeV scale leptogenesis and keV sterile neutrino DM.\\

\noindent
\textbf{\textit{Leptogenesis:}} Leptogenesis is an appealing framework to generate the observed baryon asymmetry of universe\footnote{See \cite{Buchmuller:2004nz, Davidson:2008bu} for reviews of leptogenesis.}, denoted by baryon to photon ratio $\eta_B$ defined earlier. A non-zero lepton asymmetry is first generated through lepton number (L) violating decays or scatterings which can later be converted into the observed baryon asymmetry through $(B+L)$-violating electroweak sphaleron transitions~\cite{Kuzmin:1985mm}. The sphaleron factor is given by 
\begin{equation}
    a_\text{sph}=\frac{8\,N_F+4\,N_\Phi}{22\,N_F+13\,N_\Phi}\
\end{equation}
where $N_F$ is the number of fermion generations and $N_\Phi$ is the number of Higgs doublets. For the vanilla leptogenesis scenario, we have $N_F=3\,,N_\Phi=1$ leading to $a_\text{sph} = 28/79$, requiring the lepton asymmetry at the epoch of sphaleron decoupling to be of same order as the observed baryon asymmetry.

In the absence of any fine-tuned cancellation between lepton asymmetries stored in different lepton flavours and considering only a single source of lepton asymmetry, one needs to ensure that a lepton asymmetry $\eta_{\Delta L} \sim \mathcal{O}(10^{-9})$ around the sphaleron decoupling epoch ($T_{\rm sph} \simeq 131$ GeV. In order to create a large neutrino asymmetry at later epochs, it is also necessary to ensure that the lepton asymmetry does not saturate at $T \geq T_{\rm sph}$ but continues to increase during $T_{\rm BBN} \leq T \leq T_{\rm sph}$ to $\eta_{\Delta L_\nu} \equiv \frac{\pi^2}{33\zeta(3)}\xi \sim \mathcal{O}(10^{-2})$ required to explain the Helium anomaly. Additionally, it is also important to ensure that no asymmetries are generated in charged lepton sector during this late epoch, which is ruled out from electric charge neutrality of early universe. In vanilla leptogenesis this can be naturally ensured as the charged component of the SM Higgs are no longer physical at sub-electroweak scale to appear in decay processes of type $N \rightarrow e^- H^+$. In extended and non-minimal scenarios, this can be prevented kinematically by choosing the mass spectrum of BSM particles accordingly.

Following the detailed derivation in appendix \ref{appen1}, the most general Boltzmann equations for leptogenesis from decay can be written as
\begin{align}
\frac{d\eta_N}{dz} &= -(1 + f(T)/3)\left(\tilde{n}_\gamma T\frac{\widetilde{\langle \sigma v \rangle}}{z^2\widetilde{H}} \left(\eta_N^2 - (\eta_N^{\rm eq})^2\right) \right. \nonumber \\
&+  \left.z\frac{\widetilde{\Gamma}K_1(z)}{\widetilde{H}K_2(z)} \left(\eta_N - \eta^{eq}_N\right)\right) -  \frac{f(T)}{z}\eta_N \nonumber \\
    \frac{d\eta_{\Delta L}}{dz} &= (1+f(T)/3)\left(z\epsilon_\Gamma\frac{\widetilde{\Gamma}}{\widetilde{H}}\frac{K_1(z)}{K_2(z)}(\eta_N - \eta^{\rm eq}_N)\right. \nonumber \\
    &- \left. \frac{\eta_{\Delta L}}{{\eta^{\rm eq}_l}}z\frac{\widetilde{\Gamma}}{\widetilde{H}}\frac{K_1(z)}{K_2(z)}\eta^{\rm eq}_N \right) -\frac{f(T)}{z}\eta_{\Delta L}
    \label{eq1}
\end{align}
where $\eta_N=n_N/n_\gamma, \eta_{\Delta L} = (n_L-n_{\overline{L}})/n_\gamma$ are number densities of $N, \Delta L$ normalised with respect to photon number density with $\eta^{\rm eq}_N, \eta^{\rm eq}_{l}$ being the equilibrium normalised densities. Also, in the above equations, $z = M/T$ with $M$ being the mass of the mother particle, $\widetilde{H} \equiv \sqrt{4\pi^3g_*(T)/45}M/M_{\rm Pl}$ ($g_*(T)$ is the relativistic degrees of freedom in energy density at temperature T), $\tilde{\eta}_\gamma \equiv n_\gamma/T^3= 2 \zeta(3)/\pi^2$ ($\zeta(3)=1.202$), $\widetilde{\langle \sigma v\rangle} \equiv M^2 \langle \sigma v \rangle$ (for simplicity we have taken $ \langle \sigma v \rangle \sim g^4/M^2$ with $g$ being the coupling of N with the SM bath) and $\widetilde{\Gamma} \equiv \Gamma/M$. Here $\Gamma$ denotes the total decay width of $N$, which decays into lepton and Higgs of the SM in vanilla leptogenesis scenario. The function $f(T)$, details of which can be found in appendix \ref{appen1}, is defined as $ f(T) = \frac{T}{g_{*s}}\frac{dg_{*s}}{dT}$ with $g_{*s}$ being the relativistic entropy degrees of freedom. The CP asymmetry parameter is denoted by $\epsilon_\Gamma$ while $K_i$ represents modified Bessel function of i-th order. The CP asymmetry parameter, assuming two-body decay of $N$ into lepton and Higgs doublet of the SM, is defined as
\begin{equation}
    \epsilon_\Gamma  =  \dfrac{\Gamma_{N \longrightarrow L \Phi}-\Gamma_{N \longrightarrow \Bar{L}\Phi^{*}}}{\Gamma_{N \longrightarrow L \Phi}+\Gamma_{N \longrightarrow \Bar{L}\Phi^{*}}}.
\end{equation}

Interestingly, the CP asymmetry parameter has quadratic dependence on Yukawa couplings $(y)$ even if the decay is two-body or more (i.e $\epsilon_\Gamma \sim y^2/(16\pi)$). Now assuming $y\sim 0.8$ the CP asymmetry is $\epsilon_\Gamma \sim 1\times 10^{-2}$. However, if we consider resonant regime with dominant self-energy corrections \cite{Pilaftsis:2003gt}, the Yukawa dependence of CP asymmetry can disappear allowing us to consider a large $\epsilon_\Gamma$ while varying the decay width of the mother particle independently. We consider the coupling involved in $\langle \sigma v \rangle$ to be different from Yukawa coupling, in general, in order to vary the decay width and annihilation cross-section independently.

\begin{figure}[h]
    \centering
     \includegraphics[scale=0.45]{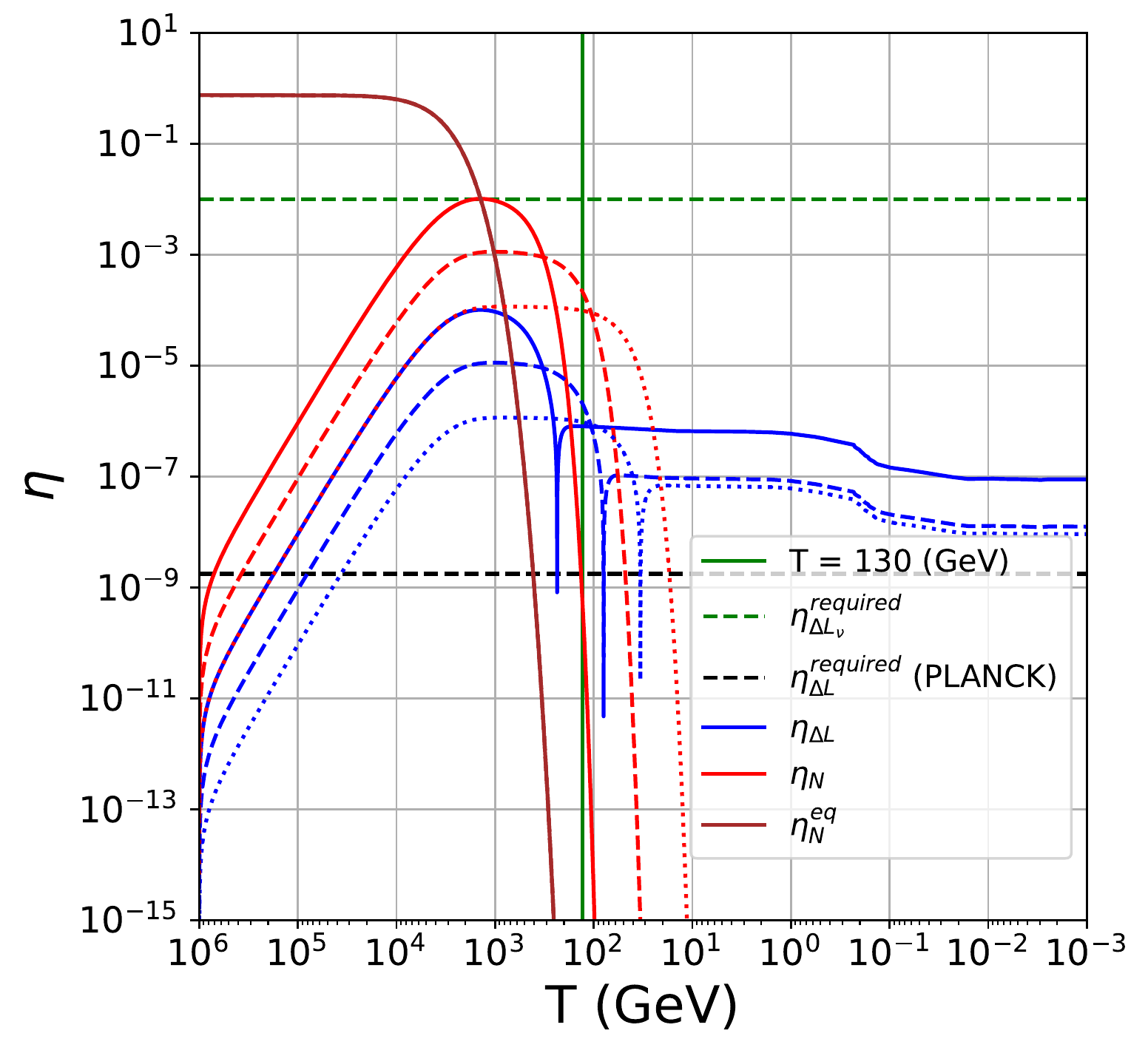} 
    \caption{Evolution for leptonic asymmetries for pure decay scenario considering benchmark points BP1 (solid), BP2 (dashed), BP3 (dash-dot) corresponding to $\tilde{\Gamma} \times M_{\rm Pl}/M = 0.1, 0.01,  0.001$ respectively with $\epsilon_\Gamma \sim 0.01$ and $M = 10$ TeV. The annihilation cross-section is assumed to be negligible.}
    \label{fig:lepto_evo1}
\end{figure}

\begin{figure}[h]
    \centering
     \includegraphics[scale=0.45]{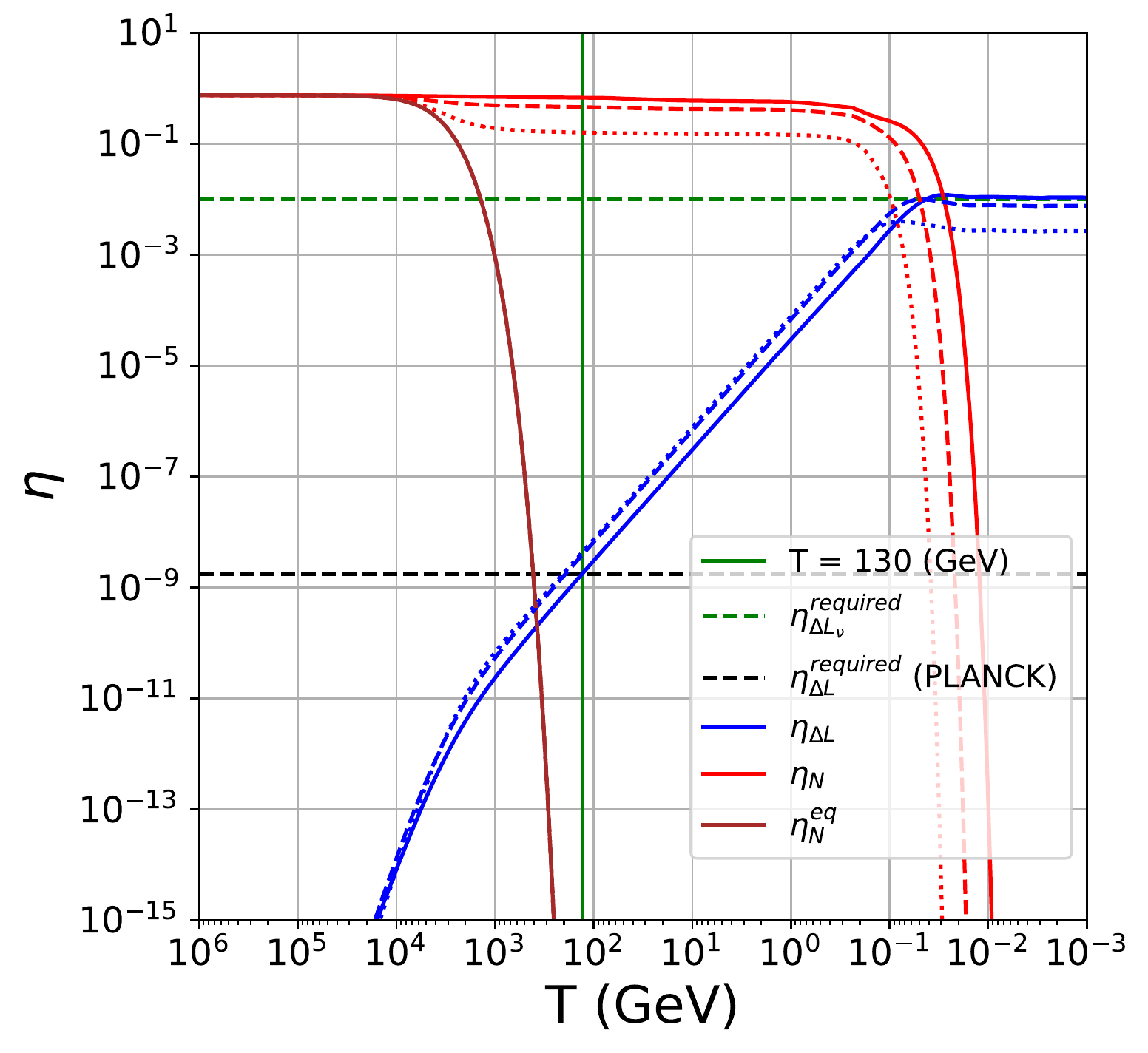}
    \caption{Evolution for leptonic asymmetries for decay scenario with non-zero additional annihilation cross-section of mother particle considering benchmark points BP4 (solid), BP5 (dashed), BP6 (dash-dot) with $\epsilon_\Gamma \sim 0.05$ and $M = 10$ TeV. The other details are shown in table \ref{tab:BP1}.}
    \label{fig:lepto_evo2}
\end{figure}

\begin{figure}[h]
    \centering
    \includegraphics[scale=0.45]{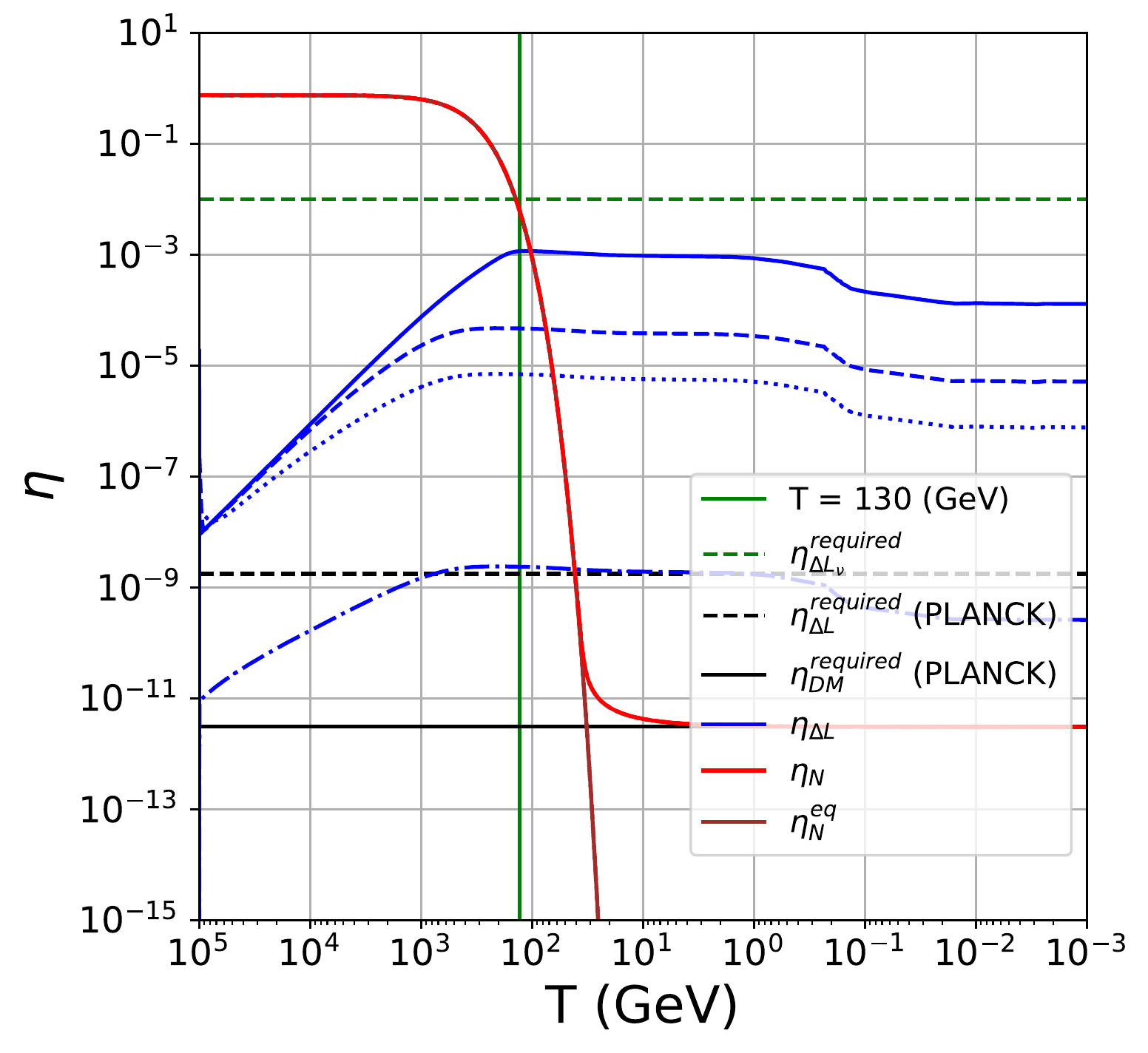}
    \caption{Evolution of leptonic asymmetries originating from $2\rightarrow 2$ processes for benchmark points BP7 (solid), BP8 (dashed), BP9 (dash-dot) with $\widetilde{\langle \sigma v\rangle}_\epsilon \times M_{\rm Pl}/M = 10^7, 10^2, 10$, $3\times 10^{-3}$ respectively with $\widetilde{\langle \sigma v\rangle}M_{\rm Pl}/M\sim 6.7\times10^{14}$, $\epsilon_\sigma \sim 0.5$ and $M = 1$ TeV.}
    \label{fig:lepto_evo3}
\end{figure}

\begin{figure*}
    \centering
    \begin{tabular}{lr}
     \includegraphics[scale=0.45]{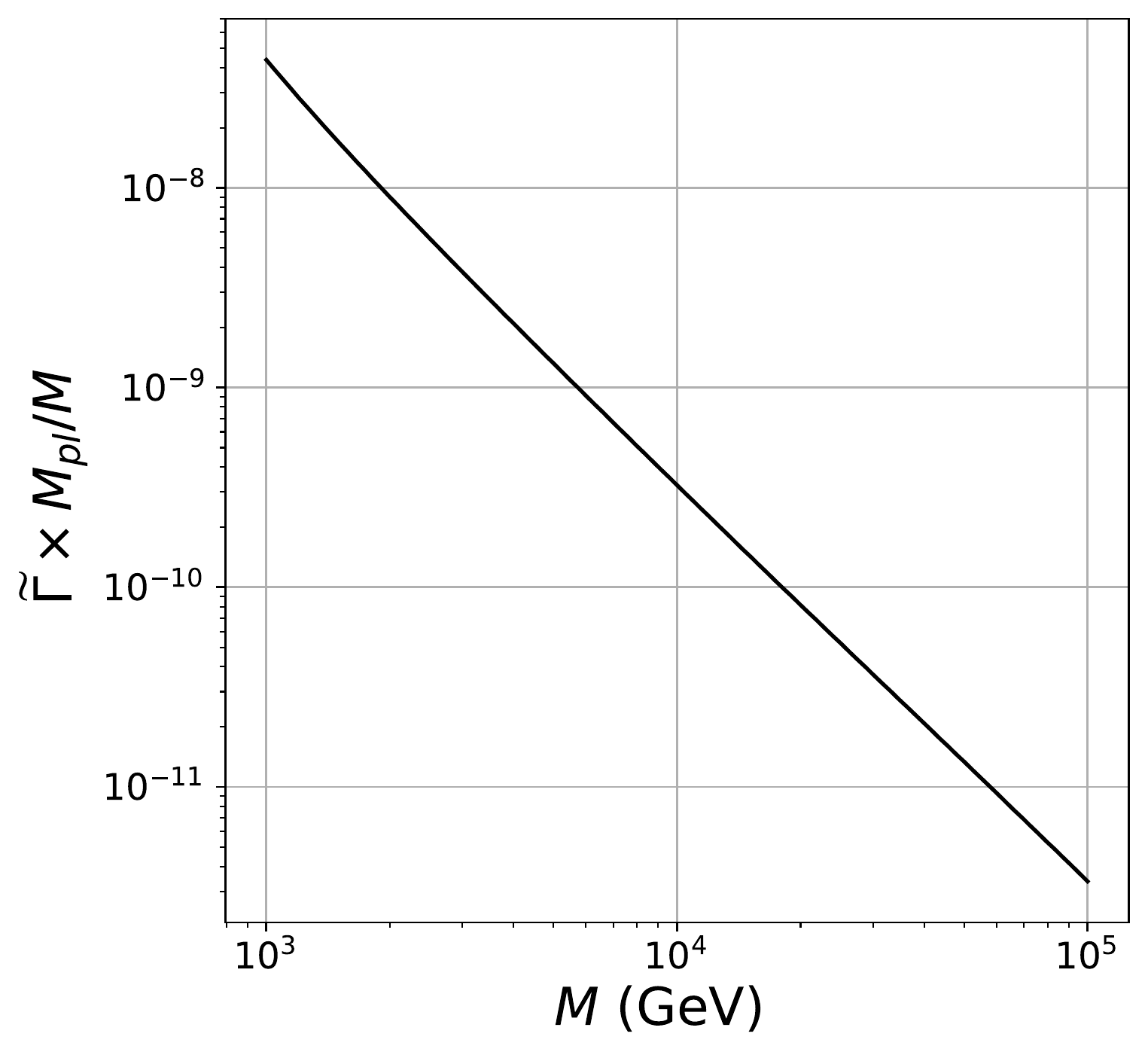}&
    \includegraphics[scale=0.45]{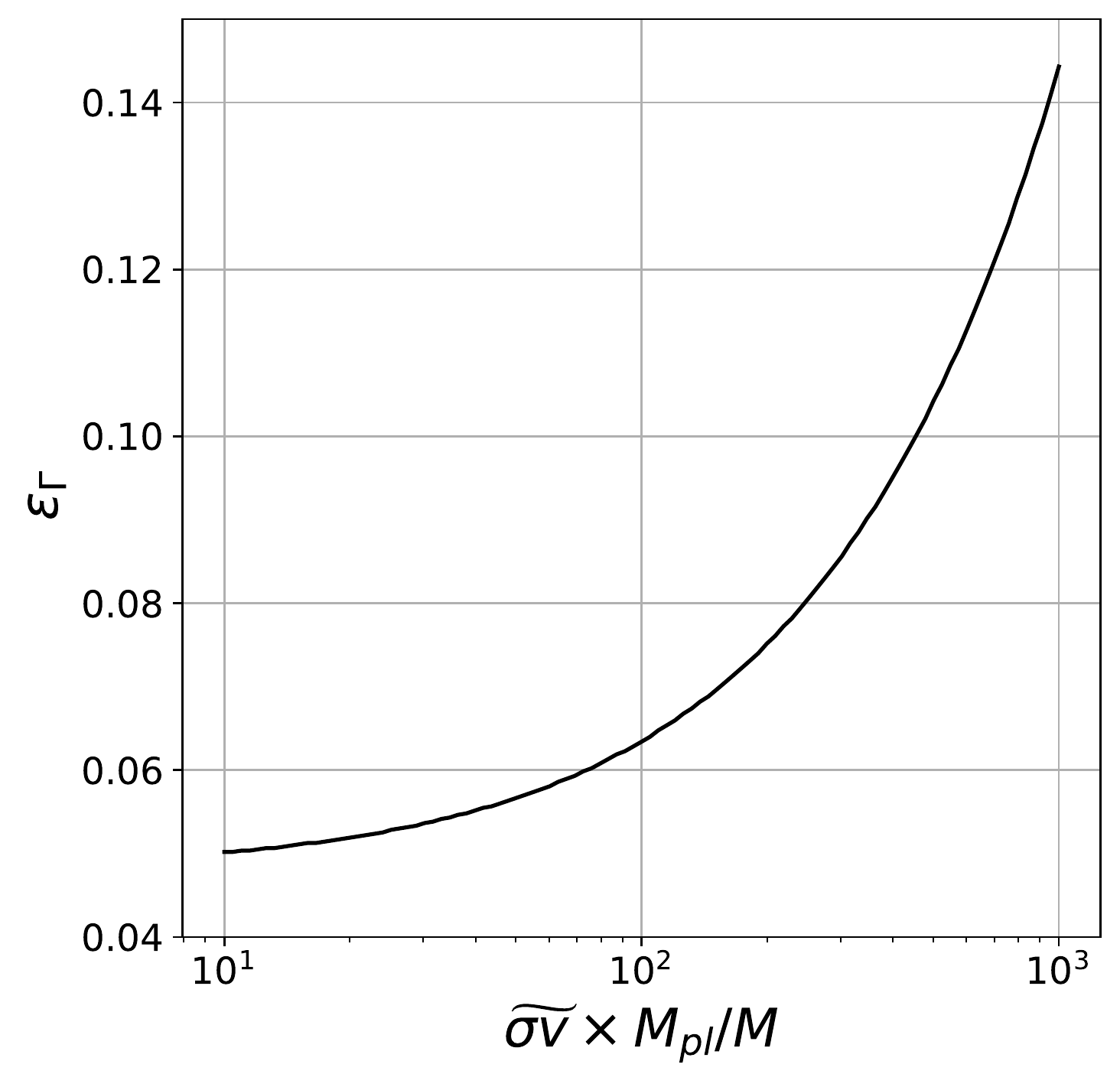}
    \end{tabular}
    \caption{Left panel: Variation of $\tilde{\Gamma}$ with the scale of leptogenesis for $\widetilde{\sigma v}M_{\rm Pl}/M=10,\,$ $\epsilon_\Gamma=0.05$. Right panel: Variation of $\epsilon_\Gamma$ with $\widetilde{\sigma v}$ for $M=1$ TeV, $\widetilde{\Gamma} M_{\rm Pl}/M = 4.4\times10^{-8}$. The choices of parameters are consistent with the required lepton asymmetry around sphaleron epoch as well as the large neutrino asymmetry at later epoch.}
    \label{fig:lepto_eps}
\end{figure*}

\begin{table}[h]
    \centering
    \begin{tabular}{|c|c|c|}
     \hline  &   $\tilde{\Gamma} \times M_{\rm Pl}/M$ &   $\widetilde{\langle \sigma v\rangle}\times M_{\rm Pl}/M$  \\
    \hline  
     BP4 & $3.1\times 10^{-10}$ & $10$ \\
     BP5  & $10^{-9}$ &  $100$ \\
     BP6  & $3.2\times 10^{-9}$ &  $1000$ \\
    \hline
    \end{tabular}
    \caption{Details of the benchmark points shown in Fig.\ref{fig:lepto_evo2}.}
    \label{tab:BP1}
\end{table}

\noindent

In Fig.\ref{fig:lepto_evo1}, we show the evolution of leptonic asymmetry for different choices of $\tilde{\Gamma} \times M_{\rm Pl}/M$ while considering resonantly enhanced and hence near maximal CP asymmetry parameter $\epsilon_\Gamma \sim \sin{2\phi}/(8\pi) \sim 0.01$ with mother particle mass $M=10$ TeV. The annihilation cross-section of the mother particle is assumed to be negligible. While we do not impose any bounds from the neutrino mass criteria applicable for specific seesaw models, the requirement of producing enhanced neutrino asymmetry below the sphaleron decoupling epoch forces the decay width $\widetilde{\Gamma}$ to be extremely small. For such small couplings, the mother particle is produced via freeze-in from the bath and decays later to produce the leptonic asymmetry. Once the decay is complete, the leptonic asymmetry gets frozen out leaving a saturated value, followed by a late dilution due to change in $g_{*s}$ (See appendix \ref{appen2} for the details of temperature variation of $g_{*s} (T)$). In this and subsequent figures, $\eta^{\rm required}_{\Delta L_\nu}, \eta^{\rm required}_{\Delta L} (\rm PLANCK)$ denote the required neutrino asymmetry and total lepton asymmetry at BBN and sphaleron decoupling epochs respectively. It can be seen from Fig. \ref{fig:lepto_evo1} that even for maximal CP asymmetry, we can not get the required leptonic asymmetry at $T=T_{\rm sph}$ while producing a large neutrino asymmetry by $T=T_{\rm BBN}$. This is due to the limited freedom one has in such a minimal setup where the decay width of the mother particle decides its production from the bath as well as the subsequent generation of lepton asymmetry. If we decouple the interactions responsible for production and decay of the mother particle, it is possible to generate the desired asymmetries simultaneously, as shown in Fig. \ref{fig:lepto_evo2}. Here, we have considered a non-vanishing annihilation cross-section of the mother particle which can lead to its production from the bath by virtue of scattering while the production from inverse decay can be sub-dominant. In such a case, the decay width can be chosen to be small in order to have a prolonged production of lepton asymmetry without compromising the production of the mother particle. Choosing the cross-section appropriately lead to early decoupling of the mother particle from the bath, leaving a large freeze-out abundance without any Boltzmann suppression. A small decay width and a large CP asymmetry can then lead to the desired lepton asymmetry at $T=T_{\rm sph}$ while producing a large neutrino asymmetry by $T=T_{\rm BBN}$, as seen from Fig. \ref{fig:lepto_evo2}. Since such a scenario can arise, in principle, for both two-body and many-body decays, we briefly comment on the viability of two-body and three-body decay leptogenesis below.

\begin{enumerate}
    \item {\it Two-body decay}: For the two-body decay, as in the vanilla leptogenesis scenarios, the reduced decay width of the mother particle depends on the Yukawa coupling as $\widetilde{\Gamma} \sim y^2/16\pi$. The decay widths chosen for benchmark points shown in Fig. \ref{fig:lepto_evo1} and Fig. \ref{fig:lepto_evo2} correspond to Yukawa couplings shown in second column of table \ref{tab:D2} assuming it to be a two-body decay. Clearly, for the benchmark points BP4-BP6 satisfying our criteria correspond to tiny Yukawa couplings. In a realistic scenario like type-I seesaw leptogenesis with RHN decaying into lepton and Higgs, this will not only create hindrance in generating the desired large CP asymmetry except in the resonant regime \cite{Pilaftsis:2003gt}, but will also be inconsistent with the neutrino mass criteria for TeV scale RHN. Therefore, the two-body decay origin of such large neutrino asymmetry consistent with successful leptogenesis is highly disfavoured, at least within simple setups.
    \begin{table}[!h]
    \centering
    \begin{tabular}{|c|c|c|}
     \hline    & $y$ (2-body) & $\Lambda$ (3-body decay, y=0.8) \\
    \hline  BP1  & $7.08\times10^{-8}$ & $9.1\times10^{6}$ GeV \\
     BP2  &  $2.2\times10^{-8}$ & $1.62\times 10^{7}$ GeV \\
     BP3  &  $7.09\times10^{-9}$ & $2.88\times 10^{7}$ GeV \\
     BP4  &  $3.9\times 10^{-12}$ & $1.22\times 10^{9}$ GeV \\
      BP5  &  $7.08\times 10^{-12}$ & $9.1\times 10^{8}$ GeV \\
       BP6  &  $1.2\times 10^{-11}$ & $6.8\times 10^{8}$ GeV \\
    \hline
    \end{tabular}
    \caption{Yukawa couplings (y) and mediator scale $\Lambda$ to get the desired decay width shown in table \ref{tab:BP1} considering two-body and three-body decay origin of leptogenesis.}
    \label{tab:D2}
\end{table}

\item {\it Three-body decay}: In the simplest realisation of three-body decay leptogenesis, the relevant process involves a mother particle decaying into lighter particles via a heavy mediator with mass denoted by a scale $\Lambda$. Assuming the Yukawa coupling involved in the two vertices involved in the decay process to be  equal $(\sim y)$, the reduced decay width
is $\widetilde{\Gamma} \sim y^4/(192\pi^3)(M/\Lambda)^4$, assuming all final state particles to be massless. The decay widths chosen for benchmark points shown in Fig. \ref{fig:lepto_evo1} and Fig. \ref{fig:lepto_evo2} correspond to mediator mass shown in the third column of table \ref{tab:D2} assuming it to be a three-body decay with Yukawa coupling $y\sim 0.8$. As we will show below, such Yukawa couplings and mediator mass can easily lead to the desired CP asymmetry without any resonant enhancement while being consistent with light neutrino masses if accommodated in a realistic seesaw scenario.

\end{enumerate}

In order to complete the discussion, we finally consider the case where the leptonic asymmetry is generated via $2 \rightarrow 2$ scattering processes \cite{Yoshimura:1978ex, Barr:1979wb, Baldes:2014gca}, similar to DM annihilations in WIMPy leptogenesis scenarios.  In such a scenario, one can satisfy all the Sakharov's conditions \cite{Sakharov:1967dj} with DM annihilations such that some of the processes responsible for WIMP freeze-out can also create a baryon or lepton asymmetry. In order to keep the washout scatterings under control, one has to ensure that the washout scatterings freeze out before WIMP freeze-out \cite{Cui:2011ab}. Adopting a model-independent approach, the Boltzmann equations for comoving number densities of the mother particle (denoted by N) and lepton number can be written as
\begin{align}
     \frac{d\eta_N}{dz} &= -(1 + f(T)/3)\left(\tilde{n}_\gamma \frac{\widetilde{\langle \sigma v \rangle}}{z^2\widetilde{H}} \left(\eta_N^2 - (\eta_N^{eq})^2\right) \right) \nonumber \\
    &-  \frac{f(T)}{z}\eta_N, \nonumber \\
    \frac{d\eta_{\Delta L}}{dz} &= (1+f(T)/3)\left(\epsilon_\sigma\frac{\tilde{n}_\gamma\widetilde{\langle \sigma v\rangle}_\epsilon}{z^2\widetilde{H}}\left(\eta^2_N - (\eta^{\rm eq}_N)^2\right)\right. \nonumber \\
    &- \left.\frac{\eta_{\Delta L}}{\eta^{\rm eq}_l} \widetilde{n}_\gamma\frac{\widetilde{\langle \sigma v \rangle}_\epsilon}{z^2 \widetilde{H}} \left(\eta^{\rm eq}_N\right)^2 \right) -\frac{f(T)}{z}\eta_{\Delta L}, 
\end{align}
where $z=M/T$ with M being the mass of the annihilating particle. The detailed derivation may be found in appendix \ref{appen1}. We denote the annihilation cross-section with leptonic final states by a subscript $\epsilon$ which together with the CP asymmetry parameter $\epsilon_\sigma$ are defined as 
\begin{align}
        \widetilde{\langle \sigma v \rangle}_\epsilon &= (\widetilde{\langle \sigma v \rangle}_L +  \widetilde{\langle \sigma v \rangle}_{\bar{L}} ); \quad
    \epsilon_\sigma = \frac{(\widetilde{\langle \sigma v \rangle}_L -  \widetilde{\langle \sigma v \rangle}_{\bar{L}})}{(\widetilde{\langle \sigma v \rangle}_L +  \widetilde{\langle \sigma v \rangle}_{\bar{L}})}.
\end{align}
Choosing some benchmark values of this CP violating cross-section, we show the evolution of N and lepton number densities in Fig. \ref{fig:lepto_evo3}. Here $\eta^{\rm required}_{\rm DM} (\rm PLANCK)$ denotes the required comoving density of DM to satisfy the relic criteria at present epoch. This can be estimated from DM density parameter at present epoch $\Omega_{\rm DM} h^2 = 0.12$ \cite{Planck:2018vyg} with $h$ being the reduced Hubble parameter. Clearly, the simplest WIMPy leptogenesis setup can not generate the required lepton asymmetry and a large neutrino asymmetry simultaneously.  

\medskip
\noindent

Therefore, only the scenario where the mother particle has a small decay width with additional production channel via scatterings and a sizeable CP asymmetry, the requirement of generating the correct lepton asymmetry by sphaleron decoupling temperature and a large neutrino asymmetry by BBN epoch can be satisfied simultaneously. We have also argued that it is more natural to fulfill these requirements in a three-body decay scenario. We now give some examples of such three-body decay model.\\

From the above model-independent discussions, we arrive at the certain constraints on decay width and CP asymmetry from the requirement of producing the correct lepton asymmetries. First of all, we notice that the dependence of the decay width $\widetilde{\Gamma}M_{\rm Pl}/M$ 
is only on the scale or the mass of the decaying particle. However, the required CP asymmetry 
$\epsilon_\Gamma$ depends upon $\widetilde{\langle \sigma v \rangle}M_{\rm Pl}/M$. When one tries to constrain the lepton asymmetry needed at the electroweak scale ($T\sim 130$ GeV) for successful leptogenesis and the late time neutrino asymmetry $\eta_{\Delta L_{\nu}} \sim 0.01$ simultaneously, we see that $\widetilde{\Gamma}M_{\rm Pl}/M$ shifts the evolution of the asymmetry horizontally i.e the freeze-out epoch of the asymmetry. Furthermore, the $\widetilde{\langle \sigma v \rangle}M_{\rm Pl}/M$ controls the final saturated value of asymmetry at freeze-out. So, to get the correct asymmetry at the electroweak scale we would need to fix the $\widetilde{\Gamma}M_{\rm Pl}/M$ for a particular mass scale $M$ and to get the correct late-time asymmetry $\eta_{\Delta L_{\nu}}$ we would need to fix the $\epsilon_\Gamma$ for a particular $\widetilde{\langle \sigma v \rangle}M_{\rm Pl}/M$. In Fig. \ref{fig:lepto_eps} we show this correlation for particular benchmark choices, consistent with the required asymmetries. While we choose a fixed value of $\widetilde{\langle \sigma v \rangle}M_{\rm Pl}/M$ for the left panel plot, we do not see much deviation from this pattern for a different choice. Similarly a different choice of $\widetilde{\Gamma}M_{\rm Pl}/M$ will not lead to a significant deviation from the pattern seen in the right panel plot. Based on this observation, we obtain the following fit for $\widetilde{\Gamma}M_{\rm Pl}/M$ and CP asymmetry parameter $\epsilon_\Gamma$ consistent with the requirements of lepton asymmetries at two different temperatures. The fit corresponding to Fig. \ref{fig:lepto_eps} can be obtained as
\begin{align}
    \widetilde{\Gamma}\times \frac{M_{\rm Pl}}{M} &=  3.20446\times 10^{-8}\left(\frac{1 {\rm TeV}}{M}\right)^2\left(1 + 0.1212\left(\frac{1 {\rm TeV}}{M}\right) \nonumber \right. \\
    &+ \left.0.243\left(\frac{1 {\rm TeV}}{M}\right)^2\right),  \\
    \epsilon_\Gamma &=  0.045974 + 9.84534\times 10^{-4} \left(\widetilde{\langle \sigma v \rangle} \frac{M_{\rm Pl}}{M}\right)^{1/2}\nonumber \\
    &+ 7.78338 \times 10^{-5}\left(\widetilde{\langle \sigma v \rangle} \frac{M_{\rm Pl}}{M}\right) \nonumber \\
    &- 1.10118\times10^{-8} \left(\widetilde{\langle \sigma v \rangle} \frac{M_{\rm Pl}}{M}\right)^{2}
\end{align}

\begin{figure}[h]
    \centering
    \begin{tabular}{lr}
     \includegraphics[trim={6.cm 22.5cm 2.cm 4.cm},clip,scale=0.9]{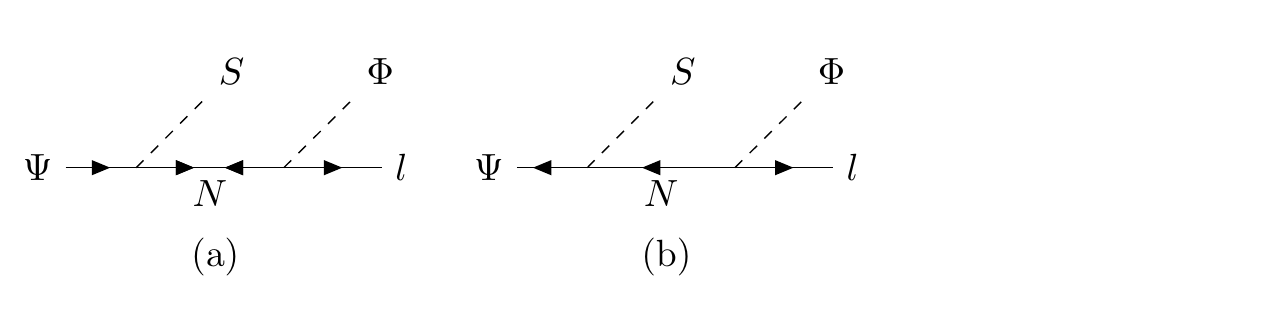}
    \end{tabular}
    \caption{Process responsible for three-body decay leptogenesis in $U(1)_{B-L}$ model.}
    \label{fig:lepto}
\end{figure}

\medskip
\noindent 
\textbf{\textit{Concrete model for 3-body decay:}} As a concrete realisation of three-body decay leptogenesis, consider an extension of the minimal gauged $B-L$ model \cite{Davidson:1978pm, Mohapatra:1980qe, Marshak:1979fm, Masiero:1982fi, Mohapatra:1982xz, Buchmuller:1991ce}. In the minimal version of this model, the SM particle content is extended with three right handed neutrinos ($N_R$) and one complex singlet scalar ($\Phi_1$) all of which are singlet under the SM gauge symmetry. The requirement of triangle anomaly cancellation
fixes the $B-L$ charge for each of the RHNs as -1. The complex singlet scalar having $B-L$ charge 2 not only leads to spontaneous breaking of gauge symmetry but also generate RHN masses dynamically. In order to realise the 3-body decay leptogenesis, we extend the model with a scalar singlets $S$ and impose a $Z_2$ symmetry under which $S$ and one of the RHNs denoted by $\psi$ are odd while all other particles are even. The usual three RHNs, even under $Z_2$ symmetry are denoted by $N_{i}$ where $i=1,2$. The relevant particle content of the model are shown in table \ref{tab:model}. The relevant part of the Yukawa Lagrangian is 
\begin{equation}
    -\mathcal{L}_Y \supset y_{l_{i \alpha}}  \overline{L}_\alpha \widetilde{\Phi} N_i + y_{\psi i} \overline{\psi^c} S N_i +{\rm h.c.}
\end{equation}
where $\Phi$ is the SM Higgs doublet. While two-body decay of the usual RHNs namely $N_i$ can lead to lepton asymmetry at high scale, there exists new source of lepton asymmetry at lower scale due to three-body decay of $\psi$, as shown in Fig. \ref{fig:lepto} using the two spinor notation \cite{Dreiner:2008tw}. The non-zero CP asymmetry will arise from the imaginary part of the resummed propagator \cite{Borah:2020ivi}. Additionally, the two RHNs $N_{1,2}$ can lead to the generation of light neutrino masses via type-I seesaw mechanism. In order to include the constraints from neutrino data we have adopted the Casas-Ibarra parametrization~\cite{Casas:2001sr} for Dirac Yukawa coupling as
\begin{align}
    y_{l_{i\alpha}} &= (\Lambda^{-1/2}\mathcal{O}\widehat{m}_{\nu}^{1/2}U^\dagger_{\rm PMNS})_{i\alpha} \, ,
\end{align}
where $\Lambda \equiv v^2/M_N$ with $M_N$ being the right handed neutrino mass and ${\cal O}$ is an arbitrary complex orthogonal matrix. Also, $\widehat{m}_\nu$ is the diagonal light neutrino mass matrix and $U_{\rm PMNS}$ is the standard Pontecorvo–Maki–Nakagawa–Sakata (PMNS) leptonic mixing matrix. Since only two RHNs take part in seesaw mechanism, we parametrize the orthogonal matrix ${\cal O}$ as
 \cite{Ibarra:2003up}
\begin{align}
{\cal O} =
\begin{pmatrix}
0 & \cos{z_{12}} & \sin{z_{12}}  \\
0 & -\sin{z_{12}} & \cos{z_{12}}
\end{pmatrix}\,,
\label{eq:rot-mat}
\end{align}


\begin{table}
\begin{center}
\begin{tabular}{|c| c | c | }
 \hline
Field & $SU(2)_{L} \times U(1)_{Y} \times U(1)_{B-L}$  & $Z_{2}$ \\
  \hline 
$N_{1,2}$ & (1, 0, -1)  &  1 \\
      $\psi$  & (1, 0, -1)     &  -1 \\ 
 $S$    & (1,0,-2)     & -1 \\
 $\Phi$ & (2, 1, 0) & 1 \\
  $\Phi_1$ & (1, 0, 2) & 1 \\
\hline
\end{tabular}
\end{center}
 \caption{Relevant particle content of the $U(1)_{B-L}$ model for three-body decay leptogenesis.}
 \label{tab:model}
\end{table}

The Yukawa coupling between $\psi$ and $N_i$ namely, $y_{\psi_i}$ in our analysis is parameterized as 
\begin{align}
    y_{\psi i} &= \rho_\psi e^{i\phi_i}
\end{align}

\begin{table}[!h]
    \centering
    \begin{tabular}{|c|c|c|c|c|c|c|c|c|}
     \hline   $m_{N_{1}}$ & $m_\Psi$ & $\rho_\psi$ & $\phi$ &$\epsilon_\Gamma$ & $z_{12}$ & $g_{B-L}$ & $M_{Z_{B-L}}$\\
    \hline  $40$ TeV & $10$ TeV & $2.3\times10^{-9}$ & $\pi/4$ & 0.021 & $4.2825\pi$i & 0.001 & 100 TeV\\
    \hline
    \end{tabular}
    \caption{Benchmark values of key parameters to get the correct lepton asymmetries from three body decay of $\psi$ in $U(1)_{B-L}$ model. The other right handed neutrino mass is fixed at $m_{N_2}=49$ TeV.}
    \label{tab:BPM}
\end{table}

\begin{figure}
    \centering
     \includegraphics[scale=0.45]{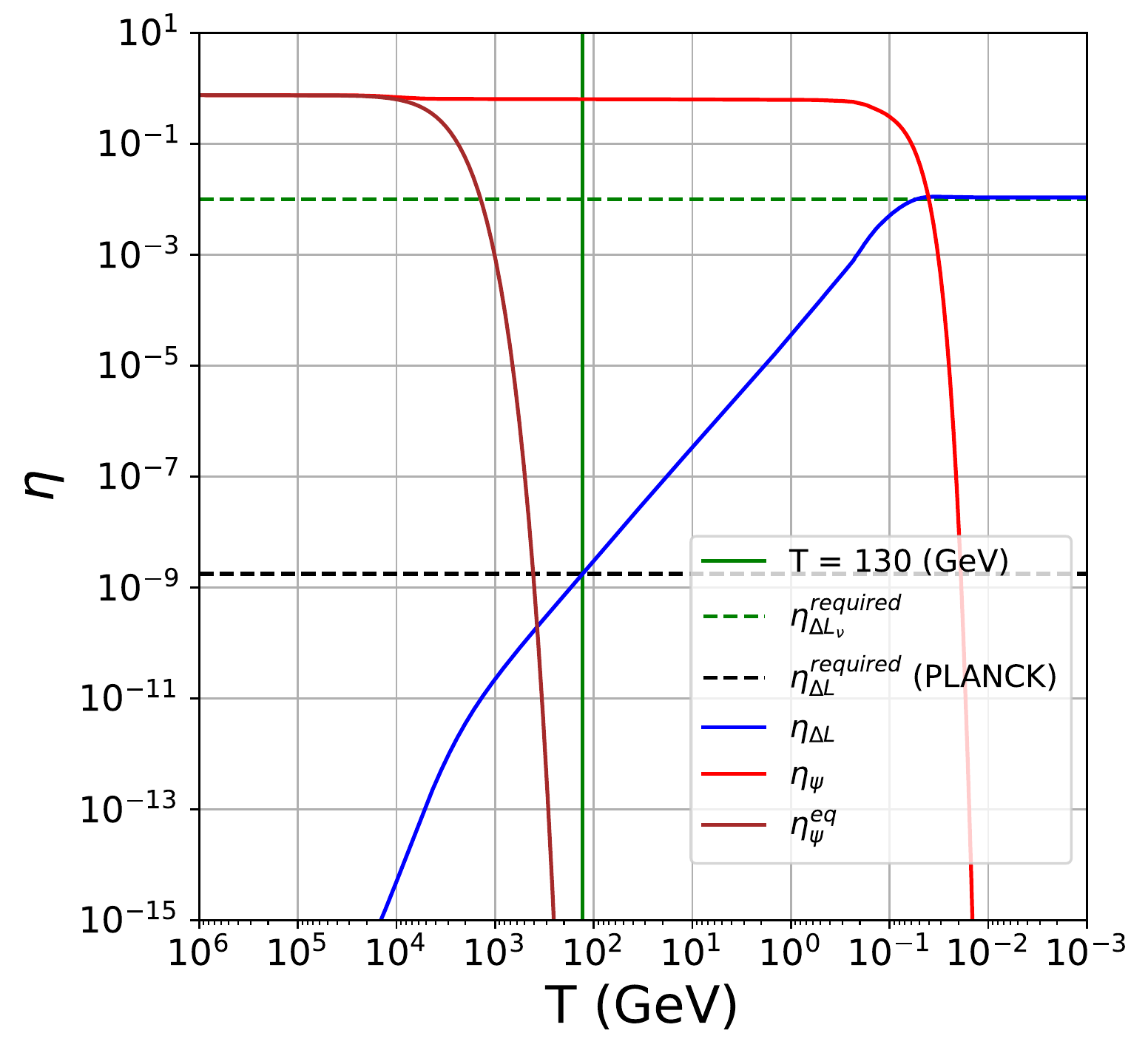}
    \caption{Evolution for leptonic asymmetries for the $U(1)_{B-L}$ model while choosing the relevant parameters as mentioned in table \ref{tab:BPM}.}
    \label{fig:lepto_evo4}
\end{figure}

The three-body decay width of $\psi$ (shown in Fig. \ref{fig:lepto}) and the corresponding $CP$ asymmetry, in the limit of massless particles apart from $N_i, \psi$ come out to be 
\begin{align}
    \Gamma(\psi\rightarrow Sl\Phi^*) &= \Gamma \simeq \sum_{i \alpha}\frac{|y_{\psi_i} y_{l_{\alpha i}}|^2}{768\pi^3}\frac{m^3_\Psi}{m^2_N} ; \\
    \Gamma_{\rm Total} &= \Gamma(\psi\rightarrow Sl\Phi^*) + \Gamma(\psi\rightarrow S^*\bar{l}\Phi) \nonumber \\
    \delta &= \Gamma(\psi\rightarrow Sl\Phi^*)-\Gamma(\psi\rightarrow S^*\bar{l}\Phi)\nonumber \\
    &=\sum_\alpha 4\frac{\Im[y_{\psi_i}y^*_{\psi_j}y^*_{l_{i\alpha}}y_{l_{j\alpha}}]}{768\pi^3}\Im[I_iI^*_j]\frac{m^3_\Psi}{m^2_N}; \nonumber \\
    &=\sum_\alpha 4\rho_\psi^2\sin(\phi_i-\phi_j)\frac{y^*_{l_{j\alpha}}y_{l_{j\alpha}}}{768\pi^3}\Im[I_iI^*_j]\frac{m^3_\Psi}{m^2_N};\nonumber \\
    \epsilon_\Gamma &= \frac{\delta}{\Gamma_{\rm Total}} \sim \sin(\phi)\sum_{k \beta} \frac{|y_{l_{\beta k}}|^2}{16\pi}\left(\frac{m_\Psi}{m_N}\right)^2. \nonumber \\
    \Im[I_j] &\simeq \sum_{\beta} \frac{|y_{l_{\beta j}}|^2}{16\pi}\left(\frac{m_\Psi}{m_N}\right)^2;\quad \Re[I_j] \simeq 1.
\end{align}
where $I_i$ is the resummed propagator of the $N_i$. While $\psi$ has Yukawa interactions with other right handed neutrinos, its dominant annihilation channel into SM particles is via $B-L$ gauge portal, the cross-section for which can be approximately written as
\begin{equation}
   \langle \sigma v \rangle \simeq \frac{g^4_{B-L}}{M^2_{Z_{B-L}}}, \end{equation}
with $g_{N-L}, m_{Z_{B-L}}$ being $U(1)_{B-L}$ gauge coupling and the mass of corresponding gauge boson respectively. Using the benchmark parameters shown in table \ref{tab:BPM}, we get 
\begin{align}
    \widetilde{\Gamma} \frac{M_{\rm Pl}}{m_\Psi} &= \frac{\Gamma}{m_\Psi} \frac{M_{\rm Pl}}{m_\Psi} \sim 7.1\times 10^{-10}, \nonumber \\
 \widetilde{\langle \sigma v \rangle}\frac{M_{\rm Pl}}{m_\Psi} &\equiv m^2_{\Psi}\langle \sigma v \rangle \left.\frac{M_{\rm Pl}}{m_\Psi}\right|_{z=1} \sim 58.7
\end{align}
which can lead to successful leptogenesis and a large neutrino asymmetry at late epochs. Note that large gauge portal interactions lead to thermal production of $\psi$ followed by its freeze-out followed by late decay into SM leptons required to produce a large neutrino asymmetry while being consistent with successful leptogenesis. We implement the model in the package \texttt{MARTY} \cite{Uhlrich:2020ltd} to extract the amplitude squared for relevant processes which are part of the Boltzmann equations. For a typical benchmark choice of parameters shown in table \ref{tab:BPM}, the corresponding evolution of $\psi$ and lepton asymmetry in terms of comoving number densities are shown in Fig. \ref{fig:lepto_evo4}. Clearly, the evolution of $\psi$ makes a departure from equilibrium at an early epoch with its slow subsequent decay leading to the generation of required lepton asymmetry by sphaleron epoch while also producing the large neutrino asymmetry at late epochs. While the model predicts vanishing lightest active neutrino mass, the $Z_2$-odd singlet scalar $S$ can be a dark matter candidate whose thermal relic can be generated due to scalar and $B-L$ gauge portal interactions.

While the particular version of $U(1)_{B-L}$ model with type-I seesaw origin of light neutrino mass discussed above is for illustrative purpose only, implementation of other seesaw mechanism is also viable.
For example, it is possible to have a scotogenic \cite{Ma:2006km} realisation of this model by considering the presence of additional $Z_2$ symmetry under which the RHNs $N_i$ and an additional scalar doublet $\eta$ are odd while all other fields are even. This allows more freedom in choosing the values of $m_N$ and $y_l$ as the radiative origin of light neutrino mass involves additional free parameters. \\

\noindent 
\textbf{\textit{Conclusion:}} Motivated by the interesting experimental signatures of a large neutrino asymmetry in the early universe, we have studied the possibility of accommodating the same in conventional leptogenesis scenarios. Considering a single source of lepton asymmetry, we have first checked if vanilla leptogenesis scenario with two-body decay can generate such asymmetries and arrived at a negative conclusion. We then consider additional interactions of the mother particle which keep it in the bath in early universe and show that for suitable scattering and decay rates with a maximal CP asymmetry parameter, it is possible to generate the required lepton asymmetry for baryogenesis and a large neutrino asymmetry at later epochs. Such a large neutrino asymmetry, via its enhancement effect on $\Delta N_{\rm eff}$ not only lead to observable signatures at CMB experiments, but can also affect BBN estimates in order to provide a solution to the recently reported Helium anomaly. We then argue in a model-independent way that the required decay width and CP asymmetry can be naturally realised in a three-body decay without requiring any resonant enhancement. Additionally, such three-body decay scenario, when implemented within a realistic seesaw model, can lead to successful prediction of light neutrino data as well whereas the simple two-body decay scenario fails to do so. We have shown a specific example of such a model by considering a gauged $B-L$ scenario. With growing evidences suggesting the presence of large neutrino asymmetry around the BBN epoch will lead to stronger hints at a TeV scale leptogenesis scenario with three-body decay origin of lepton asymmetry. This is also interesting in view of the tight constraints  on high scale leptogenesis scenarios which create large individual lepton flavour asymmetries while keeping the total lepton asymmetry small, for successful leptogenesis. Such constraints arise due to the generation of helical hypermagnetic field which can source a new contribution to the baryon asymmetry of the universe, as pointed out recently in \cite{Domcke:2022uue}. Since only low scale leptogenesis scenarios are capable of producing such large neutrino asymmetry at late epochs while being consistent with the correct baryon asymmetry of the universe, other laboratory based experiments can offer complementary probes keeping it verifiable in near future. Future CMB experiments like CMB-S4 \cite{Abazajian:2019eic} with sensitivity upto $\Delta {\rm N}_{\rm eff}
= 0.06$ will also be able to constrain the neutrino asymmetry significantly, keeping the verifiability of our setup promising. Converting this sensitivity to neutrino asymmetry gives $\zeta_\alpha \sim 0.37$ with $\eta_\nu = 0.1$. A large neutrino asymmetry in the early universe can also reduce the Hubble tension \cite{Barenboim:2016lxv}. Additionally, from a model building perspective, if dark matter in the universe is composed of keV sterile neutrino, the proposed scenario can offer a DM-baryon cogenesis setup by generating the required neutrino asymmetry for resonant production of DM \cite{Shi:1998km}. \\

\acknowledgements
DB would like to acknowledge the hospitality at PITT-PACC, University of Pittsburgh during May 2022 where this work was initiated.

\appendix

\section{Derivation for the Boltzmann Equations}
\label{appen1}
We start with the general Boltzmann equation for a massive particle whose number can change due to decay as well as $2 \leftrightarrow 2$ annihilations into standard model bath particles. In terns of number density, this can be written as \cite{Kolb:1990vq}

\begin{align}
    \frac{d{n}_F}{dt} + 3Hn_F &= -\langle \sigma v\rangle \left(n^2_F - (n^{\rm eq}_F)^2\right) - \Gamma_D (n_F - n^{\rm eq}_F).\label{eq:A1}
    \end{align}

Similarly, the most general equation for asymmetry number density $n_{\Delta f} = n_f-n_{\overline{f}}$ considering it to be generated from both annihilation and decay of heavy particle $F$ can be written as
    \begin{align}
    \frac{d{n}_{\Delta f}}{dt} + 3Hn_{\Delta f} &= \epsilon_\sigma \langle \sigma v\rangle \left(n^2_F - (n^{\rm eq}_F)^2\right) + \epsilon_\Gamma \Gamma_D (n_F - n^{\rm eq}_F) \nonumber \\
    &- n_{\Delta f}\frac{n^{\rm eq}_F}{n^{\rm eq}_f}\left(\Gamma_D + \langle \sigma v \rangle n^{\rm eq}_F \right),
    \label{eq:A2}
\end{align}
where $n_f$ is the number density, $\langle \sigma v \rangle$ is the thermal-averaged cross-section \cite{Gondolo:1990dk} and $\Gamma_D=\Gamma K_1(M/T)/K_2(M/T)$ is the thermal-averaged decay width \cite{Kolb:1979qa, Buchmuller:2004nz}. $H$ is the Hubble parameter related to energy density via Friedman's equation as 
$ H^2 = \frac{8\pi G}{3} \rho$. Using $\rho=\frac{\pi^2}{30} g_*(T) T^4$ in a radiation dominated universe, we get 
$$ H = \sqrt{\frac{4\pi^3G g_*(T)}{45}}T^2 = \sqrt{\frac{4\pi^3 g_*(T)}{45}}\frac{T^2}{M_{\rm Pl}}.$$

In order to absolve the expansion of the Universe we start with the invariance of the total entropy of the Universe $S=sa^3$ 
\begin{align}
    \frac{dS}{dt} &= \frac{ds}{dt}a^3 + 3a^2s\frac{da}{dt} = 0, \nonumber \\
    \Dot{s} &= -3Hs \quad \Dot{s} = \frac{ds}{dt}. 
    \label{eq:A3}
\end{align}
where $s=\frac{2\pi^2}{45}g_{*s} T^3$ and $a$ are the entropy density and the scale factor respectively. Now, defining comoving number density $Y_F = n_F/s$, the LHS of Eq. \eqref{eq:A1} will be re-written as
\begin{align}
    \Dot{Y}_F &= \frac{\Dot{n}_F}{s} - \frac{n_F}{s^2}\Dot{s}, \nonumber \\
    s\Dot{Y}_F &= \Dot{n}_F + 3Hn_F. \label{eq:A4}
\end{align}
In order to change the variable from $t$ to $z=M/T$ we start from Eq. \eqref{eq:A2}
\begin{align}
    \frac{dT}{dt}\frac{ds}{dT} &= -3Hs, \nonumber \\
    \frac{dT}{dt}\left(\frac{3s}{T} + \frac{s}{g_{*s}}\frac{dg_{*s}}{dT}\right) &= -3Hs, \nonumber \\
    \frac{dT}{dt} &= -\left(1 + \frac{T}{3g_{*s}}\frac{dg_{*s}}{dT}\right)^{-1}HT, \nonumber \\
    \Dot{Y}_F &= -\left(1 + \frac{T}{3g_{*s}}\frac{dg_{*s}}{dT}\right)^{-1}HT\frac{dY_F}{dT}, \nonumber \\
    \Dot{Y}_F &= -\frac{1}{(1+f(T)/3)}HT\frac{dY_F}{dT}, \nonumber \\
    f(T) &= \frac{T}{g_{*s}}\frac{dg_{*s}}{dT}.\label{eq:A5}
\end{align}
Considering $z=M/T$ as the variable leads to $\frac{dY_F}{dT}=-\frac{dY_F}{dz} \frac{M}{T^2}$. Using these in Eq. \eqref{eq:A1} and ignoring the second term on the RHS leads to the well-known equation \cite{Edsjo:1997bg}
\begin{equation}
    \frac{dY_F}{dz} = -\sqrt{\frac{\pi}{45 G}} \frac{h^{1/2}_* M}{z^2} \langle \sigma v \rangle \left ( Y^2_F - (Y^{\rm eq}_F)^2 \right ),
\end{equation}
where 
$$ h^{1/2}_* = \frac{g_{*s}}{\sqrt{g_*}} (1+f(T)/3).$$
In order to rewrite the equations in terms of $\eta_F = n_F/n_\gamma$, we make the following rearrangements.
\begin{align}
    s\Dot{Y}_F &= -s\frac{HT}{(1+f(T)/3)}\frac{d}{dT}\left( \frac{\widetilde{n}_\gamma}{\widetilde{s}}\eta_F\right) \nonumber \\
    &= n_\gamma \frac{H}{(1+f(T)/3)}\left(\frac{T}{g_{*s}}\frac{dg_{*s}}{dT}\right)\eta_F - n_\gamma \frac{HT}{(1+f(T)/3)}\frac{d\eta_F}{dT} \nonumber \\
    s\Dot{Y}_F &= n_\gamma \frac{Hf(T)}{(1 + f(T)/3)}\eta_F - n_\gamma \frac{HT}{(1+f(T)/3)}\frac{d\eta_F}{dT}
\end{align}
where $\widetilde{\langle \sigma v \rangle} = M^2 \langle \sigma v \rangle , \widetilde{\Gamma} = \Gamma/M$,  $\tilde{s} = s/T^3$ and $\tilde{n}_\gamma = n_\gamma/T^3$. This is same as the first equation in Eq. \eqref{eq1}.

Now, using this in Eq. \eqref{eq:A1} and Eq. \eqref{eq:A4}, we have 
\begin{widetext}
\begin{align}
    n_\gamma \frac{Hf(T)}{(1 + f(T)/3)}\eta_F - n_\gamma \frac{HT}{(1 + f(T)/3)}\frac{d\eta_F}{dT} &= -n^2_\gamma \langle \sigma v \rangle (\eta_F^2 - (\eta_F^{\rm eq})^2) - n_\gamma \Gamma_D (\eta_F - \eta^{\rm eq}_F) \nonumber \\
    - n_\gamma \frac{HT}{(1 + f(T)/3)}\frac{d\eta_F}{dT} &= -n^2_\gamma \langle \sigma v \rangle (\eta_F^2 - (\eta_F^{\rm eq})^2) - n_\gamma \Gamma_D (\eta_F - \eta^{\rm eq}_F) - n_\gamma \frac{Hf(T)}{(1 + f(T)/3)}\eta_F \nonumber \\
    \frac{\widetilde{H}T}{(1 + f(T)/3)}\frac{d\eta_F}{dz} &= -\tilde{n}_\gamma T\frac{\widetilde{\langle \sigma v \rangle}}{z^2} \left(\eta_F^2 - (\eta_F^{\rm eq})^2\right) -  \Gamma_D \left(\eta_F - \eta^{\rm eq}_F\right) -  \frac{Hf(T)}{(1 + f(T)/3)}\eta_F \nonumber \\
    \frac{d\eta_F}{dz} &= -(1 + f(T)/3)\left(\tilde{n}_\gamma T\frac{\widetilde{\langle \sigma v \rangle}}{z^2\widetilde{H}} \left(\eta_F^2 - (\eta_F^{\rm eq})^2\right) +  z\frac{\widetilde{\Gamma}K_1(z)}{\widetilde{H}K_2(z)} \left(\eta_F - \eta^{\rm eq}_F\right)\right) \nonumber \\
    &-  \frac{f(T)}{z}\eta_F
\end{align}
\end{widetext}
where $\widetilde{H} = HM/T^2$ and $z=M/T$.
Now, moving over to the Boltzmann equation for the asymmetry \eqref{eq:A2}
\begin{align}
    \frac{d\eta_{\Delta f}}{dz} &= (1+f(T)/3)\left(\epsilon_\sigma\tilde{n}_\gamma\frac{\widetilde{\langle \sigma v\rangle}}{z^2\widetilde{H}}\left(\eta^2_F - (\eta^{\rm eq}_F)^2\right) \right. \nonumber \\
    &+\left. z\epsilon_\Gamma\frac{\widetilde{\Gamma}}{\widetilde{H}}\frac{K_1(z)}{K_2(z)}(\eta_F - \eta^{\rm eq}_F)\right. \nonumber \\
    &- \left. \frac{\eta_{\Delta f}}{\eta^{\rm eq}_f} \left(z\frac{\widetilde{\Gamma}}{\widetilde{H}}\frac{K_1(z)}{K_2(z)}\eta^{\rm eq}_F + \tilde{n}_\gamma\frac{\widetilde{\langle \sigma v\rangle}}{z^2\widetilde{H}} (\eta^{\rm eq}_F)^2 \right)\right) -\frac{f(T)}{z}\eta_{\Delta f}
\end{align}

\section{Temperature variation of $g_*, g_{*s}$}
\label{appen2}

The effective relativistic degrees of freedom contributing to the energy density of the universe is defined as 
\begin{equation}
    g_*(T) = \sum_{i \in {\rm boson}} g_i \left ( \frac{T_i}{T} \right)^4 + \frac{7}{8} \sum_{i \in {\rm fermion}} g_i \left ( \frac{T_i}{T} \right)^4.
\end{equation}
Similarly, the relativistic entropy degrees of freedom is defined as 
\begin{equation}
    g_{*s}(T) = \sum_{i \in {\rm boson}} g_i \left ( \frac{T_i}{T} \right)^3 + \frac{7}{8} \sum_{i \in {\rm fermion}} g_i \left ( \frac{T_i}{T} \right)^3.
\end{equation}
Here, only the particles with mass $m_i \ll T$ contributes. Also $T_i=T$ when the particle species i (with internal degrees of freedom $g_i$) is in equilibrium with the photon bath. At very high temperature $T>200$ GeV, when all the standard model particles are relativistic, $g_*=106.75$ is maximum. Also, at high temperatures $g_{*}=g_{*s}$. Only after neutrino decoupling around $T \sim \mathcal{O}(2 \, \rm MeV)$ leading to different neutrino and photon temperatures $T_\gamma/T_\nu = (11/4)^{1/3}$, we have $g_{*} \neq g_{*s}$ with the latter being slightly larger. We start with $g_*=g_{*s}=106.75$ at $T>200$ GeV and calculate them at lower temperatures by removing the contribution of those standard model particles which become non-relativistic, simply by using the criteria $m_i > T$. As the temperature falls, heavy particles becomes non-relativistic and stop contributing to $g_*(T), g_{*s}(T)$. Below the QCD phase transition temperature $ T \sim \mathcal{O}(200 \, \rm MeV)$, we drop the contributions of light quarks and gluons to $g_*(T), g_{*s}(T)$ due to the formation of hadronic bound states. At temperature below 100 MeV, only electrons, neutrinos and photons contribute to $g_*(T), g_{*s}(T)$. Finally, at even lower temperatures, below MeV (after neutrino decoupling and $e^- \, e^+$ annihilation), only neutrinos and photons contribute to relativistic degrees of freedom keeping $g_{*s} > g_*$ as mentioned earlier.

\end{document}